\documentclass[a4paper]{article}
\usepackage[a4paper]{geometry}
\usepackage{graphicx}
\usepackage{authblk}

\newcommand{\dsfrac}[2]{\displaystyle{\frac{#1}{#2}}}
\def\simless{\mathbin{\lower 3pt\hbox
{$\rlap{\raise 5pt\hbox{$\char'074$}}\mathchar"7218$}}}   
\def\simmore{\mathbin{\lower 3pt\hbox
{$\rlap{\raise 5pt\hbox{$\char'076$}}\mathchar"7218$}}}   

\def\difd{\mathrm{d}}

\title{Numerical simulations of the jetted tidal disruption event Swift J1644+57}

\begin{document}
\author{Petar Mimica}
\author{Miguel A. Aloy}
\affil{Departamento de Astronom\'ia y Astrof\'isica, Universidad de Valencia, 46100, Burjassot, Spain}
\author{Dimitrios Giannios}
\affil{Department of Physics and Astronomy, Purdue University, 525 Northwestern venue, West Lafayette, IN 47907, USA}
\author{Brian D. Metzger}
\affil{Columbia Astrophysics Laboratory, Columbia University, New York, NY, 10027, USA}
\maketitle

\begin{abstract}
  In this work we focus on the technical details of the numerical
  simulations of the non-thermal transient Swift J1644+57, whose
  emission is probably produced by a two-component jet powered by a
  tidal disruption event. In this context we provide details of the
  coupling between the relativistic hydrodynamic simulations and the
  radiative transfer code. First, we consider the technical demands of
  one-dimensional simulations of a fast relativistic jet, and show to
  what extent (for the same physical parameters of the model) do the
  computed light curves depend on the numerical parameters of the
  different codes employed. In the second part we explain the
  difficulties of computing light curves from axisymmetric two
  dimensonal simulations and discuss a procedure that yields an
  acceptable tradeoff between the computational cost and the quality
  of the results.
\end{abstract}

\section{Introduction}\label{sec:intro}

Tidal disruption events (TDE) of stars by massive black holes
\cite{Lacy:1982,Rees:1988,Evans:1989} provide one of the ways to
detect an inactive massive black hole in the center of a galaxy. Stars
whose orbit brings them too close to a black hole can be disrupted and
a large fraction of their mass can be accreted. This sudden accretion
event is expected to produce a luminous UV and X-ray TDE flare
\cite{Strubbe:2009,Lodato:2011}. It has recently been argued that TDEs
can also produce jets
\cite{Giannios:2011,Krolik:2012,Tchekhovskoy:2014}. These jets are
expected to be detectable due to their non-thermal emission, and in
radio they are likely to appear as transients that peak on the
timescales of about a year \cite{Giannios:2011,Metzger:2015} after the
tidal disruption.

Swift J1644+57 (SwJ1644 in the rest of the article) is a non-thermal
transient detected on March 28th.\,2011
\cite{Bloom:2011,Burrows:2011,Levan:2011}. It is thought to be
prototype of a jetted TDE (jTDE). Its radio emission
\cite{Zauderer:2011,Berger:2012,Zauderer:2013} is long-lived and
complex. One of the first attempts to explain the radio emission of
Swift J1644+57 using one-dimensional models and relativistic
hydrodynamic simulations\cite{Metzger:2012} assumed that the emission
is coming from the shocks that form when the jet interacts with the
circumnuclear medium (CNM). This model is more analogous to a GRB
afterglow than to an AGN\footnote{Interested readers can
  find details of the model based on the AGN jet-disc symbiosis in
  Section~2.1 of \cite{vanVelzen:2011}. We note that
  \cite{vanVelzen:2011} assume that the origin of the emission is
  \emph{internal} to the jet, as opposed to the \emph{external}
  emission that we advocate.}. While the model was successful in
explaining the first few weeks of SwJ1644 radio emission, it turns out
that the long-term emission is not easily accounted for by a radio
afterglow from a single ultrarelativistic, narrow jet. The reason for
this is that the source rebrightened after $\simmore 150$ days, the
time at which the afterglow model predicted a steady
decline\cite[Figure~1]{Berger:2012}. The late time rebrightening
implies that additional energy is injected in the external shock
months after the burst\cite{Berger:2012}. This may happen if, for
instance, the jTDE is accompanied by a powerful mildly relativistic
component\cite{Mimica:2015}\footnote{For an overview of alternative
  explanations for late-time rebrightening see Section~2.2 of
  \cite{Mimica:2015}.}.

In this contribution we focus on the numerical and technical aspects
of jTDE modeling. In Section~\ref{sec:num} we describe the
numerical codes used in the calculations and how they depend on each
other. The Section~\ref{sec:sim} gives an overview of computational
requirements and their consequences on the kind of models that we were
able to produce. The discussion and summary is given in the Section~\ref{sec:sum}.

\section{Numerical codes}\label{sec:num}

SwJ1644 is a distant object (its redsfhit is $z = 0.354$
\cite{Bloom:2011}), and is thus too small to be resolved in great
detail\footnote{According to our numerical models we predict that
  SwJ1644 may become resolvable by Very Long Baseline Inteferometry
  sometime in the next few years (see Figure~13 in
  \cite{Mimica:2015}).}. Thus, the only direct observations that we
have are the total fluxes for different times and frequencies (i.e.,
multiwavelength light curves). Nevertheless, when simulating this
source we need to construct a detailed, resolved model of the SwJ1644
jet, simulate its dynamics, compute its resolved emission, and only in
the final stage integrate the resolved emission to compute the total
flux to be compared to observations. In order to achieve this we use
two numerical codes so that the output of the first code (\emph{MRGENESIS}) is used as
an input to the second one (\emph{SPEV}). The former simulates the jet
dynamics, while the latter computes its emission.

\subsection{Relativistic hydrodynamics simulations}\label{sec:hydro}

\emph{MRGENESIS} is a code developed to solve the equations of
relativistic numerical (magneto) hydrodynamics {\em in the laboratory
  frame (i.e., a frame at rest with respect to the {\em distant
    source})}\footnote{Interested readers can find details about the
  original \emph{GENESIS} in \cite{Aloy:1999}, its extension to
  relativistic magnetohydrodynamics in \cite{Leismann:2005}, and the
  current parallelized version (\emph{MRGENESIS}) that handles
  ultrarelativistic magnetized outflows in \cite{Mimica:2009}.}. The
code is massively parallelized using MPI\footnote{MPI (Message Passing
  Interface) is used for distributed-memory parallelism, see
  http://www.mpi-forum.org} and OpenMP\footnote{OpenMP (Open
  Multi-Processing) is used for shared-memory parallelism, see
  http://openmp.org.} librararies, and uses the parallel
HDF5\footnote{HDF5 (Hierarchical Data Format) is used to store and
  manage large datasets, see http://hdfgroup.org.}  library for file
input and output. We employ a method of lines for our
  simulations in which the spatial reconstruction is a based on the third order PPM
reconstruction routines\cite{Colella:1984}, and the time
  update is performed by means of a third order TVD Runge-Kutta
scheme\cite{Shu:1988}. Numerical fluxes are evaluated with the Marquina flux
formula\cite{Donat:1998}. We simulate SwJ1644 in two dimensions
assuming axisymmetry and discretize the computational domain using
spherical coordinates. The black hole is assumed to be located
at the origin and the TDE occurs at scales that are much smaller than
the ones we consider in our work (we simulate neither the TDE
nor the jet launching). We set the beginning of the numerical grid at
$\sim 10^{16}$ cm. The jet is injected through the inner boundary in
an angular interval $[0, \theta_j]$, assuming a constant Lorentz
factor $\Gamma_j$, and a luminosity whose time dependence is
$L_j(t) = L_{j,0}\max[1, (t/t_0)]^{-5/3}$. $t_0$ is the time during
which the jet luminosity is constant, and the peak lumiosity $L_{j,0}$
is computed from the total jTDE energy $E_{\rm ISO}$ through
$L_{j,0} = 2 E_{\rm ISO}/ (5 t_0)$. The jet interacts with a
CNM that is assumed to be at rest with a density
profile $\rho_{\rm cnm} = n_{18}m_p (r/10^{18}{\rm cm})^{-k}$, where
$m_p$ is the proton mass, $n_{18}$ is the particle number density at
$10^{18}$ cm and $k$ is the power-law index of the density profile. We
assume that initially both the jet and the external medium are
cold. During the simulation we periodically store the snapshots of the
computational grid in order to obtain a sufficient coverage of the
jet-CNM interaction to be able to compute the resulting emission with
sufficient precision (see Section~\ref{sec:sim}).

\subsection{Radiative transfer}\label{sec:rad}

Once the relativistic hydrodynamics simulation has finished, the
snapshots it produced are used by the code \emph{SPEV} to compute the
emission.  The spatial distribution of the hydrodynamics variables
needs to be stored with sufficient frequency because of the space-time
transformations that need to be performed to compute the
observable radiation in the (distant) {\em observer
  frame}. \emph{SPEV}\cite{Mimica:2009b} is designed to process an
arbitrary number of pre-computed sets of hydrodynamic states of the
fluid (i.e., density, pressure, positions, velocities, magnetic field)
and to compute the resulting non-thermal emission. The procedure can
be summarized as follows:
\begin{enumerate}
\item {\bf Injection}: \emph{SPEV} uses a set of Lagrangian particles
  (LPs) as a representation of the emitting volumes. Each LP has a
  defined position, shape and velocity. Attached to it there is a
  representation of the non-thermal electron (NTE) energy
  distribution. In \emph{SPEV}, new LPs are created each time a
  snapshot is processed. The position of the shock fronts that form as
  a consequence of the jet-CNM interaction is detected, and LPs are
  injected into the shocked fluid immediately behind the front. Since
  the actual process of particle acceleration occurs on temporal and
  spatial scales far below the ones we are able to simulate (see e.g.,
  \cite{Sironi:2015} for a recent review), we use a phenomenological
  model to link the fluid state with NTE distribution. The model we
  use is a standard one \cite{Sari:1998}: the energy contained in the
  NTEs and in the stochastic magnetic field are assumed to be
  fractions $\epsilon_e$ and $\epsilon_B$ of the shocked fluid thermal
  energy. The injection spectrum is assumed to be a power-law in
  energy $n(\gamma) \propto \gamma^{-p}$, where $p$ is the power-law
  index and $\gamma$ is the NTE Lorentz factor. We determine the
  normalization and energy cutoffs of the distribution using a
  numerical procedure described in detail in Sections 3.2 and 3.3 of
  \cite{Mimica:2012}.
\item {\bf Transport, evolution and emission}: As it processes the
  snapshots, \emph{SPEV} transports the injected LPs, assuming that
  after they have been injected they move with the fluid. An exception
  are the newly injected LPs: between the snapshot in which the
  particle is injected and the subsequent one the two edges parallel
  to the shock front are assumed to move with different velocities
  (one with the velocity of the shocked fluid and the other one with
  the faster velocity of the shock front). This ensures that new LPs
  correctly represent the increase in the volume due to newly shocked
  fluid, regardless of the numerical resolution or the frequency with
  which snapshots are stored. Simultaneously, the NTE energy losses
  due to synchrotron radiation, as well as the energy gains(losses)
  due to adiabatic compression(expansion) are taken into account:
\begin{equation}\label{eq:losses}
  \dsfrac{\difd \gamma}{\difd t} = \dsfrac{1}{3}\dsfrac{\difd\ln
    \rho}{\difd t}\gamma -
  \dsfrac{4}{3}\dsfrac{c\sigma_T}{m_ec^2}\left(u_B+u_{\rm rad}\right)\gamma^2\,
\end{equation}
where $t$, $\rho$, $u_B$ and $u_{\rm rad}$ are the time, fluid
density, magnetic field energy density and external radiation field
energy density (all measured in the LP frame). Section~3.2 of
\cite{Mimica:2009b} describes an efficient numerical algorithm for
solving Eq.~\ref{eq:losses} for each NTE distribution in each LP. At
the same time that the particles are evolved, their non-thermal
emission is computed (see Section~4 of \cite{Mimica:2009b} for a
description of an efficient numerical algorithm for computing
synchrotron emission and absorption coefficients, and Section~A.2.1 of
the same paper for a description of the way particles in
two-dimensional axisymmetric models can be used to represent
three-dimensional emitting volumes). The coefficients computed in this
step are stored in a global array that is used in the final step.
\item {\bf Solving the radiative transfer equation}: After all the
  snapshots have been processed and the emission from all LPs has been
  computed for all times, the time- and frequency-dependent emission
  can be computed. We assume that there is a virtual screen (``virtual
  detector'', VD) located in front of the source, and the goal is to
  compute the intensity in each of the virtual pixels of the VD. This
  is done by solving the radiative transfer equation
\begin{equation}\label{eq:intens}
  \dsfrac{\difd I_\nu}{\difd s} =j_\nu - \alpha_\nu I_\nu, ,
\end{equation}
where $I_\nu$, $j_\nu$ and $\alpha_\nu$ are the specific intensity,
emissivity and absorption coefficient. $s$ is the distance along the line
of sight from the jet to a particular pixel and it relates the
simulation time $t$ to the time of observation $T$ by $t = T + (D -
s)/c$, where $D$ is the distance from the VD to the source. In other
words, for a given $T$ and $t$, $s$ is the distance of a LP from which
radiation is observed at time $T$ if that LP is emitting at a time
$t$. We note that Eq.~\ref{eq:intens} needs to be solved in such a way
that $s$ is monotonously increasing from the jet towards the
VD. Therefore, after all the snapshots have been processed, the
emission and absorption coefficients need to be sorted according to
their $s$ and then Eq.~\ref{eq:intens} is solved for each pixel.
\end{enumerate}
In practice the positions of LPs in steps (i) and (ii) are stored in a
set of (large) intermediate files (``preprocessor'' files). A
postprocessor executes the radiation part of step (ii) by reading the
preprocessor files and storing the coefficients in memory. After all
preprocessor files have been postprocessed, the step (iii) is
executed. Both preprocessor and postprocessor parts of \emph{SPEV} are
paralellized using OpenMP. The postprocessor, in particular, needs to
be run on a machine with enough shared memory\footnote{Interested
  readers should consult Chapter 4 of \cite{Aloy:2013} for a detailed
  description of the optimized memory management in SPEV.} because all
information produced in step 2 needs to be held in memory for the step
3 to correctly compute the intensities.

\section{Simulations of Swift J1644+57 jet}\label{sec:sim}

As mentioned in Section~\ref{sec:hydro}, we model SwJ1644 jet in
two-dimensions. However, as will be explained in this Section, the
total cost in terms of computing time, disk storage and working memory
of each two-dimensional simulation is high. Therefore, we first
perform a number of much less costly tests using a one-dimensional
approximation to two-dimensional models. The goal of those tests is to
determine the values of a number of the parameters of the numerical
simulation for which the results are satisfactory. Afterward, the 
two-dimensional simulations are performed using the values determined
in one-dimensional tests. 

\subsection{One-dimensional tests}\label{sec:1D}

Using \emph{MRGENESIS} we simulate one-dimensional spherically
symmetric jets. For the purpose of computing their emission with
\emph{SPEV} we assume that their opening angles are $\theta_j$. We fix
all the physical parameters of the jet
($E_{\rm ISO} = 4\times 10^{54}$ erg, $\Gamma_j = 10$,
$\theta_j = 0.1$ rad, $t_0 = 5\times 10^5$ s) and of the external
medium ($n_{18} = 60$ cm$^{-3}$, $k = 1.5$). This corresponds to the
inner relativistic component of the two-component jet in
\cite{Mimica:2015}. The numerical parameters that will be varied
are\footnote{We note that the results of a limited number of numerical
  tests have also been presented in Appendix~A of \cite{Mimica:2015},
  but here we perform an exhaustive analaysis.}:
\begin{itemize}
  \item $\Delta T_{\rm snap}$: time interval between snapshots
    (\emph{MRGENESIS})
  \item $\Delta r$: radial resolution (\emph{MRGENESIS})
  \item $\Delta\theta$: angular resolution
    (\emph{MRGENESIS}\footnote{When performing one-dimensional simulations we
      create mock two-dimensional snapshots using $\Delta \theta$ as
      angular cell size.})
  \item $N_b$: number of NTE energy distribution bins (\emph{SPEV})
  \item $n_x$: resolution of the virtual detector (\emph{SPEV}\footnote{In these
      one-dimensional tests we assume that the jet is observed on-axis (viewing angle
      is $0$ degrees), so that its emission is axially-symmetric and
      we only need to compute the emission along VD x-axis. For
      off-axis observations a (much more costly) two-dimensional VD
      calculation is performed \cite[Section~5.2]{Mimica:2015}. We
      comment in detail on technical challenges of 2D simulations in
      next section.})
\end{itemize}

We choose as a default set of parameters
$\Delta T_{\rm snap} = 8.3\times 10^4$ s, $\Delta r = 5\times 10^{13}$
cm, $\Delta \theta =2\times 10^{-3}$ rad,
$N_b = 32$, $n_x = 1000$. We follow the jet evolution until
$\sim 6.7\times 10^8$ seconds. During the first $1.3 \times 10^8$
seconds the snapshots are output every $\Delta T_{\rm snap}$, and
afterward the time interval increases by a factor $1.002$ after every snapshot
is output.

\begin{figure}
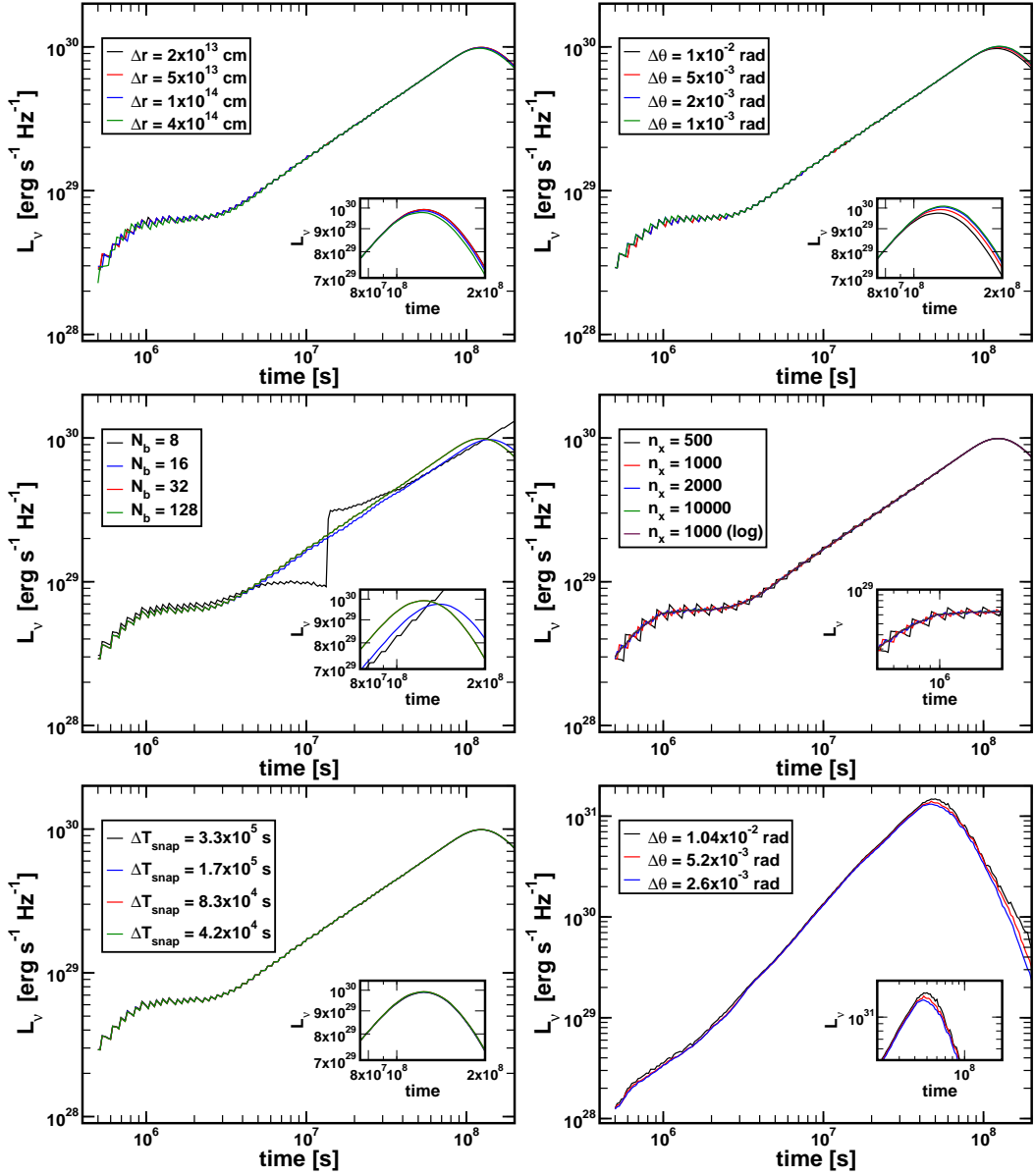
\label{fig:var1D}
\begin{center}
\includegraphics[scale=0.25]{figures/vardr.eps}\hspace{0.1cm}\includegraphics[scale=0.25]{figures/varntheta.eps}
\includegraphics[scale=0.25]{figures/varnb.eps}\hspace{0.1cm}\includegraphics[scale=0.25]{figures/varnx.eps}
\includegraphics[scale=0.25]{figures/varTsnap.eps}\hspace{0.1cm}\includegraphics[scale=0.25]{figures/lc2D.eps}
\end{center}
\caption{$1.4$ GHz radio light curves of one-dimensional tests
  (Section~\ref{sec:1D}). The default model is shown in red. The
  insets show the zoom-in around the light curve maximum. \emph{Top
    left panel}: Variations in $\Delta r$. \emph{Top right panel}:
  Variations in $\Delta\theta$. \emph{Middle left panel}: Variations in
the number of bins $N_b$. Note that $N_b=32$ and $N_b=128$ curves
overlap. \emph{Middle right panel}: Variations in the number of
VD pixels. Note that the curves for $n_x=10^4$ and $n_x=10^3$
with a logarithmic distribution of pixels overlap. \emph{Bottom left
  panel:} Variations in $\Delta T_{\rm snap}$. \emph{Bottom right
  panel:} Emission from two dimensional two-component jet for
different effective angular resolutions (see Section~\ref{sec:2D}).}
\end{figure}

Fig.~\ref{fig:var1D} shows the light curves for the
default model evaluated at 1.4\,GHz (red line) and the consequence of
variations in the resolution of the hydrodynamic models, i.e., as a
result of changing $\Delta r$. As can be seen, relatively small
  differences of the order of a percent are visible at early times
and after the light curve maximum. The top right panel in
Fig.~\ref{fig:var1D} shows the result of changing angular
resolution. Here the differences are somewhat larger, especially when
comparing the lowest and highest resolution ($\sim 4$\% at peak). The
middle left panel in Fig.~\ref{fig:var1D} shows how the results depend
on the number of bins in which NTE energy distribution of each LP is
discretized. We see that the convergence is only achieved for
$N_b\simmore 32$. In the middle right panel we can see that the
oscillations at early times are due to poor VD resolution. The angular
size of the jet is very small at those times and it is only properly
resolved for $n_x\sim 10^4$ for a linear pixel distribution (covering
the interval [$0$, $5\times 10^{18}$ cm]), and $n_x\sim 10^3$ for a
logarithmic one (covering the interval [$10^{12}$ cm,
$5\times 10^{18}$ cm]). In the next section we will explain how we
solved the problem of noise for 2D simulation light curves, especially
the off-axis ones where it is not feasible to distribute the VD pixels
logarithmically. Finally, the bottom left panel shows the (weak)
dependence on $\Delta T_{\rm snap}$.

\subsection{Two-dimensional simulations}\label{sec:2D}

In the previous section we showed how the light curves change as the
five principal numerical parameters are changed. Most importantly for
the 2D simulations, with the previous tests we have already obtained a
very good proxy for the largest values of $\Delta r$ and
$\Delta \theta$ we can use in order to reduce the computational
cost without compromising the accuracy of the results. In the
final simulations showed in \cite{Mimica:2015} we used
$\Delta r = 4\times 10^{14}$ cm and
$\Delta \theta \simeq 2.6\times 10^{-3} $ rad. In this section we
discuss the technical challenges of computing the emission even from
these relatively low-resolution hydrodynamic simulations.

The complete two-dimensional evolution is stored in $2412$
snapshots. It was fortunate that HDF5 format permits data compression
so that the compressed snapshots occupy $\simeq 460$ Gbytes. Running
the preprocessor with the default parameters of Section~\ref{sec:1D}
produces $160$ (compressed) intermediate files. Each of the first
$159$ files contains the positions, geometry, velocity, magnetic field
and the $32$ bins of the NT energy distibution of $2.5\times 10^6$
LPs, while the last file only contains $\simeq 1.9\times 10^6$
LPs. This means that in total there are $\simeq 4\times 10^8$
two-dimensional emitting LPs in step (ii) of the \emph{SPEV} algorithm
(see Section~\ref{sec:rad}). Since each of the LPs has to be converted
to a three-dimensional emitting volume, it may contribute to multiple
VD pixels and also to multiple observing times, we soon realized that
the realistic upper limit of $\simless 1$ TByte of RAM on our shared
memory machine\footnote{We use the machine \emph{LluisVives} hosted
  at the University of Valencia. It has 30 Xeon 7500 hexacore CPUs
  clocked at 2.67 GHz and almost 1 TBytes of RAM. For more
  information see
  http://www.uv.es/siuv/cas/zcalculo/calculouv/des\_vives.wiki (in
  Spanish).} would not be enough to compute off-axis images at full
resolution. A remedy to this problem was to artificially reduce the
angular resolution when preprocessing the hydrodynamic snapshots by a
factor of $2$ or $4$ (this would roughly correspond going from the
second-best to third- of fourth-best angular resolution in top left
panel in Figure~\ref{fig:var1D}.). This reduces the number of emitting
volumes so that even off-axis radiative transfer calculations could be
performed.

\subsubsection{On-axis light curves}\label{sec:onaxis}
For the purposes of the numerical tests we computed the on-axis light
curves (in complete analogy to the tests presented in
Section~\ref{sec:1D}). Bottom right panel in Figure~\ref{fig:var1D}
shows on-axis light curves using the full angular resolution, as well
as the result of degrading it by factors of 2 and 4. 

\begin{figure}
\begin{center}
\includegraphics[scale=0.35]{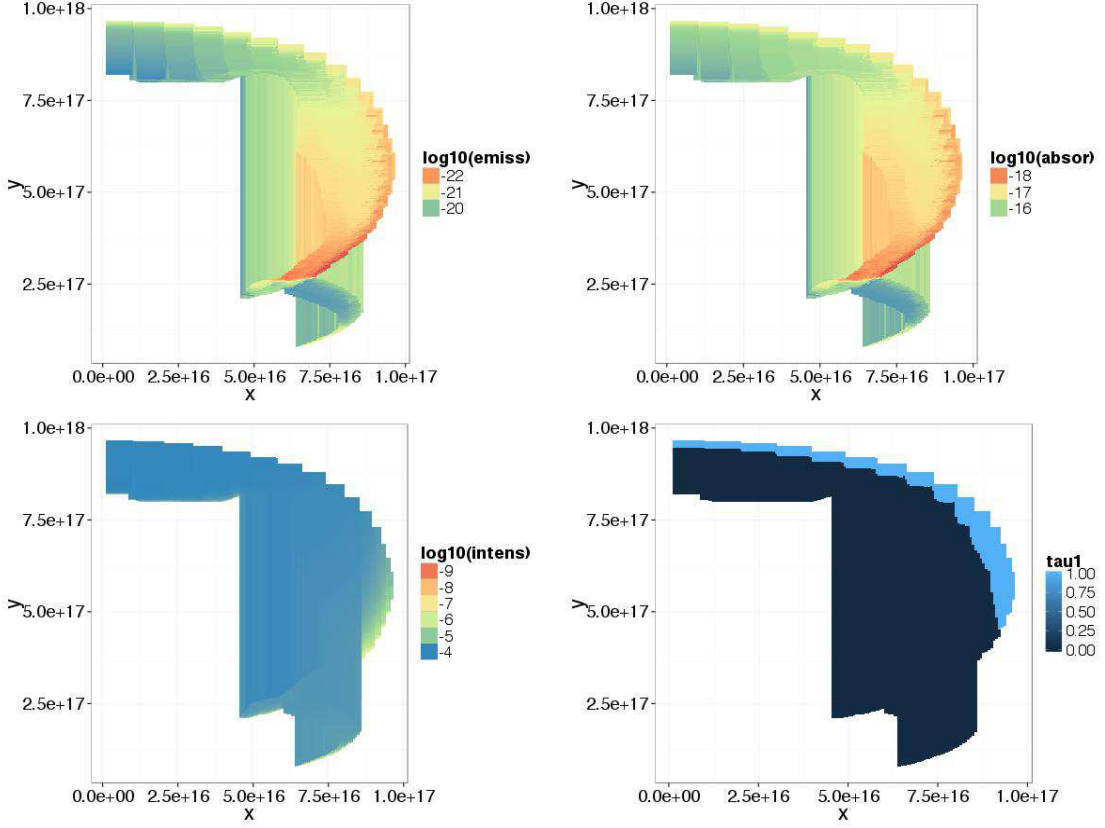}
\end{center}
\caption{Observer frame simultaneous emission, absorption, intensity
  and optical depth that contribute to the emission at $T=1.4\times 10^6$
  seconds at $15.4$ GHz. \emph{Top left}: synchrotron emissivity in erg s$^{-1}$
  cm$^{-3}$ Hz$^{-1}$. \emph{Top right}: synchrotron self-absorption
  in cm$^{-1}$. \emph{Bottom left}: specific intensity in erg s$^{-1}$
  cm$^{-2}$ Hz$^{-1}$. The intensity is the result of integrating
  Equation~\ref{eq:intens} from bottom to top for each vertical line.
  \emph{Bottom right}: optical depth measured from the observer
  (integral of the absorption from top to bottom along vertical lines). The
  light blue colored region denotes the part of the emitting volume
  that is transparent, while the emission in the dark blue region is
  absorbed. As explained in Section~\ref{sec:onaxis}, the information
  displayed here is a subset of the data stored in the global
  array used by \emph{SPEV} to compute the intensity in each pixel
  (see Section~\ref{sec:rad}). The luminosity is obtained by
  integrating the intensity along vertical lines in the top right panel
  (taking into account that the emission is axially symmetric).}
\label{fig:observer}
\end{figure}

In Figure~\ref{fig:observer} we show a subset of the data that
\emph{SPEV} stores when postprocessing the intermediate files. In this
particular case we show all the contributions to the emission at
$T=1.4\times 10^{6}$ seconds and at 15.4\,GHz. The data shown
originates from LPs emitting at different times from different
positions (the lower the value of $y$ the earlier the LP had to emit
for it to be observed at $1.4\times 10^{6}$ seconds). In other words,
what is shown does not correspond to one particular snapshot of the jet
evolution. Nevertheless, we know that the edge of the emitting region
corresponds to the position of the bow shock of the jet. We also see
in the bottom right panel in Figure~\ref{fig:observer} that a small
region behind the shock is optically thin and most of the contribution
to the emission comes from this region and from the surface at which
the emission becomes optically thin (the transition between light and
dark blue regions). Using this type of analysis it is
  possible to understand, among other things, why the inverse Compton
cooling of the electrons is not very efficient at low frequencies
(though it may be important for higher frequency observations of
SwJ1644, see \cite{Kumar:2013}): the electrons cool after being
accelerated by the shock, but since they are advected behind
it, once they enter the opaque region (dark blue), they stop
contributing to the observed emission. Therefore the observer does not
see a large effect of the inverse Compton cooling
\cite[Section~4.1.2]{Mimica:2015}.

\subsubsection{Off-axis light curves} Here we comment on the procedure
we use to reduce the numerical noise that shows at early times
(Figure~\ref{fig:var1D}). As was commented in Section~\ref{sec:1D},
the jet is small at early times, so that at a typical resolution of
$1000$ points per spatial direction it is very poorly resolved by the
VD. Due to memory limitations we were not able to use VD resolution of
$10000^2$. However, the resolutions of $1000^2$ pixels were
feasible. Therefore, we adopted the following procedure:
\begin{enumerate}
  \item Pre-determine the jet viewing angle $\theta_{\rm obs}$ and the times $T_i$ ($i =
    1, \ ..\ n_t$) at which it is to be observed.
  \item Compute the radio images using $1000\times 1000$ VD for each
    $T_i$ in a single calculation.
  \item For each $T_i$, determine the rectangle that contains $99.99\%$ of the
    emission. Then expand the rectangle by a safety margin.
  \item For each $T_i$, independently re-compute the radio image by
    limiting the VD to the rectangle computed in the previous step,
    but using the resolution $1000\times 1000$.
\end{enumerate} 
The above procedure enables us to compute the emission from early
times without numerical artifacts\footnote{Figures~12-14 in
  \cite{Mimica:2015} were computed using this procedure. Note that in
  Figure~14 in \cite{Mimica:2015} the jet size in the upper panels is
  $\simeq 3-5$ times smaller than in the lower panels, but all are
  computed using the same VD resolution.}. The tradeoff is that we
need to postprocess the intermediate files multiple times: once in
step (ii) and then additional $n_t$ times in step (iv). The computing
time does not necessarily scale as $n_t$ since, for each $T_i$, the
emission and absorption coefficients are only computed from those
emitting volumes that contribute to the emission, while the rest
(typically corresponding to most of the volume of the model) are
ignored. This is made efficient by indexing the intermediate
files. For a given observing angle the calculations at $8$ different
observing frequencies last $\simeq 3-5$ days.

\section{Summary}\label{sec:sum}

In this contribution we have discussed the technical details of the
one and two dimensional simulations of the SwJ1644 jet. In
Section~\ref{sec:num} we explain in detail how the relativistic
hydrodynamic simulations are coupled to the radiative transfer code to
produce the synthetic observations (light curves) that can be directly
compared to the observations. In Section~\ref{sec:1D} we show that
many one-dimensional tests can be used to probe the sensitivity of the
computed light curves on the resolution of the hydrodynamic
simulations, on the frequency of stored snapshots, on the resolution
of the discretization in energy space and of the virtual detector. The
results obtained helped to determine the resolution of the two
dimensional simulation so that its computational cost is acceptable,
but also so that the final light curves are as free of numerical
artifacts as possible. This procedure has been successful in enabling
us to produce a viable two-component jet model of the SwJ1644 jet.

\section*{Acknowledgements}
PM and MAA acknowledge the support from the European Research Council
(grant CAMAP-259276), and the partial support of grants
AYA2013-40979-P, PROMETEO-II-2014-069 and SAF2013-49284-EXP .  DG
acknowledges support from the NASA grant NNX13AP13G.  BDM acknowledges
support from the NSF grant AST-1410950 and the Alfred P. Sloan
Foundation. We thankfully acknowledge the computer resources,
technical expertise and assistance provided by the "Centre de C\`alcul
de la Universitat de Val\`encia" through the use of {\emph{Llu\'{\i}s
    Vives}} cluster and {\emph{Tirant}}, the local node of the Spanish
Supercomputation Network. We use the open source software packages
\emph{Grace}\footnote{http://plasma-gate.weizmann.ac.il/Grace} and
\emph{ggplot2}\footnote{http://ggplot2.org} to produce the graphics in
this paper.

\end{document}